

 \input harvmac

\def\pyidk{PHY-9057135}
\def\CA{{\cal A}} \def\CB{{\cal B}}

  \def\CO{{\cal O}}

%
%
%
%
%

%

\def\bar#1{\overline{#1}}
\def\vev#1{\left\langle #1 \right\rangle}

%

\def\frac#1#2{{\textstyle{#1\over #2}}} 
%
%
%
%

\def\GeV{{\rm GeV}}

%
%
%
%

%
%
\def\ltap{\ \raise.3ex\hbox{$<$\kern-.75em\lower1ex\hbox{$\sim$}}\ }
\def\gtap{\ \raise.3ex\hbox{$>$\kern-.75em\lower1ex\hbox{$\sim$}}\ }
\def\gl{\ \raise.5ex\hbox{$>$}\kern-.8em\lower.5ex\hbox{$<$}\ }
\def\roughly#1{\raise.3ex\hbox{$#1$\kern-.75em\lower1ex\hbox{$\sim$}}}
%
%
        
\def\eg{\hbox{\it e.g.}}

\def\np#1#2#3{{Nucl. Phys. } B{#1} (#2) #3}
\def\pl#1#2#3{{Phys. Lett. } {#1}B (#2) #3}

\def\physrev#1#2#3{{Phys. Rev. } {#1} (#2) #3}

\relax

\def\frac#1#2{{\textstyle{#1 \over #2}}}

\def\[{\left[}
\def\]{\right]}
\def\({\left(}
\def\){\right)}

\def\sc{\sin\theta_c}
\def\cc{\cos\theta_c}
\def\tc{\tan\theta_c}
\def\db{\Delta\CB}
\def\sb{\Delta\CB'}
\def\mw{m_{\tilde w}}
\def\mus{m_{\tilde u}}
\def\mds{m_{\tilde d}}
\def\pdi{\db_{\{1,i\}}}
\def\pdii{\db_{\{2,i\}}}

\noblackbox

\Title{\vbox{
\hfill NSF-ITP-94-69,\  LBL-35807  \smallskip
\hfill DOE/ER/40561-148-INT94-00-61\vskip -.6in}}
{Family
Symmetries and Proton Decay$^*$
\footnote{}{$^*$ This work was supported
in part by the Director, Office of Energy
Research, Office of High Energy and Nuclear Physics, Division of High
Energy Physics of the U.S. Department of Energy under Contract
DE-AC03-76SF00098, by DoE grant DOE-ER-40561,
and by NSF grants PHY-90-57135 and PHY-89-04035.} }
\vskip -.4in
\centerline{Hitoshi Murayama\footnote{$^a$}{On leave of absence
from Department of Physics, Tohoku University, Sendai, 980 Japan. Email:
{\tt murayama@lbl.gov}}}
\centerline{{\sl Lawrence Berkeley Laboratory}}
\centerline{{\sl Berkeley, CA 94720, USA}}
 \medskip\centerline{{\it and}}\medskip
\centerline{D. B. Kaplan\footnote{$^b$}{Address before 7/1/94: Institute for
Theoretical Physics, UC Santa Barbara, Santa Barbara CA, USA 93106-4030.
 Email:
{\tt dbkaplan@ben.npl.washington.edu}}}
 \centerline{{\sl
Institute for
Nuclear Theory HN-12}}
\centerline{{\sl University of Washington}}
\centerline{{\sl  Seattle WA 98195, USA}}

 \bigskip\baselineskip 14pt
 \noindent

{The proton decay modes $p\to K^0 e^+$ and $p\to K^0 \mu^+$ may be
visible in certain supersymmetric theories, and if seen would provide
evidence for new flavor physics at extremely short distances.  These
decay modes can arise from the dimension five operator
$(Q_1Q_1Q_2L_{1,2})$, where $Q_i$ and $L_i$ are ${i^{th}}$ generation
quark and lepton superfields respectively. Such an operator is not
generated at observable levels due to gauge or Higgs boson exchange in a
minimal GUT.  However in theories that explain the fermion mass
hierarchy, it may be generated at the
Planck scale with a strength such that the decays $p\to K^0 \ell^+$ are
both compatible with the proton lifetime and visible at
Super-Kamiokande.  Observable proton decay can even occur in theories
without unification.}
\Date{6/94}
\baselineskip 18pt
\newsec{Introduction}
There is an $SU(3)^5$ chiral flavor symmetry in the standard model in the
limit that the Yukawa couplings vanish.  It is possible that the Yukawa
couplings are themselves the fundamental parameters that break these
symmetries, but their hierarchical structure suggests that they are
fossils left over from a simpler form of flavor symmetry violation at
short distances. It is therefore interesting to examine flavor changing
processes for effects that would not occur if the Yukawa couplings were
the sole source of flavor violation.  Since the Yukawa couplings of the
first two families are very small, flavor changing
processes involving the first two families are sensitive tests
for the existence of new flavor physics at some scale $\Lambda$ that
violates the first family chiral symmetries directly. In the
standard model, flavor changing
operators are suppressed by either $1/\Lambda$ in the case of neutrino
masses, or $1/\Lambda^2$ for four fermion operators leading to $B$
violation, rare decays and flavor changing neutral currents (FCNC).  In
the supersymmetric standard model the power counting is different
 due to the existence of squarks.  There are new sources of
flavor symmetry violation including dimension two squark masses, and
both dimension four and five $B$ and $L$ violating operators.  The first
two sources are problematic for supersymmetry and must be eliminated;
the dimension four $B$ and $L$ violating operators may be forbidden by
imposing a symmetry, such as either $R$-parity or a flavor symmetry,
while the lack of observed FCNC suggests the existence of some kind of flavor
symmetry for the squark mass matrix.

In this Letter we focus on the dimension five $B$ and $L$ violating
operators in supersymmetric (SUSY) theories, which allow one to examine flavor
physics at extremely short
distances.  It may not be obvious why such operators contain information
about flavor.
The effective $B$-violating operators in the original
(non-supersymmetric) grand unified theories (GUTs) are
four-fermi operators of quark and lepton fields induced by an exchange
of GUT-heavy gauge boson  (for a review, see \ref\langacker{P.~Langacker, Phys.
Rept. 72 (1981) 185.}). Therefore the structure of the operators are
determined solely by the gauge quantum numbers of the fields under the
unified group. An observation of proton decay, such as the mode $p
\rightarrow \pi^0 e^+$, would reveal the gauge group structure
at extremely high energy scale.

In SUSY models, on the other hand, $B$-violating operators arise due to
the flavor structure of the models rather than the gauge structure.
 Supersymmetric $B$-violating dimension four and five
operators are necessarily flavor-off-diagonal due to the Bose symmetry
of the superfields. For example, the dimension four operator
$\epsilon^{abc} U^c_a D^c_b D^c_c$ exists only when two down quark $D^c$
fields belong to different generations. The same is true for
dimension five operators such as $\epsilon_{abc} \epsilon^{\alpha\beta}
\epsilon^{\gamma\delta} Q_\alpha^a Q_\beta^b Q_\gamma^c L_\delta$ which
will be the main topic in this Letter. Therefore $B$-violating operators
are presumably intimately related to the physics which generate Yukawa
interactions and flavor mixings. Indeed, the dimension five operators
are generated in generic SUSY-GUT models via the exchange of
color-triplet Higgses, and the coupling constants are those of Yukawa
interactions.

Because $B$ violation can occur through such low dimension
operators, as opposed to dimension 6 in original GUTs,
SUSY models are sensitive to flavor physics through $B$ violation all the
way up to the Planck scale.
For instance, the operator ${1\over M_p} (Q_2 Q_2)(Q_1 L_i)$ gives
a proton lifetime shorter than the experimental bound by about 14 orders of
magnitude. Since proton decay is a flavor changing process, further
suppression can arise if the operators are suppressed by Yukawa
couplings, as is the case when they are induced by Higgs triplet
exchange in a GUT.  As we discuss below, when the sole source of flavor
symmetry breaking are the Yukawa couplings themselves, then the decays
$p\to K^0 \ell^+$ will be unobservable, being suppressed by the $u$
quark Yukawa coupling. Alternatively, flavor symmetries broken
at a scale $\Lambda$ can give a suppression of the form $\Lambda/M_p$ to
some power which can both explain the long lifetime of the proton, as
well as render the decay modes $p\to K^0 \ell^+$ observable at
Super-Kamiokande --- even in string-inspired models without gauge
unification below the string scale.

In the next section, we briefly review the form of dimension five $B$
violating operators in supersymmetric theories, and discuss their flavor
structures. We then show that $p \rightarrow K^0 \ell^+$ modes are
unobservable in conventional GUT-models. Therefore, an observation of
these modes would signal a different structure of flavor physics at very
short distances. We point out in \S 4 that
operators suppressed by powers of Planck mass can
significantly contribute to $p\rightarrow K^0 \ell^+$ modes
with branching fractions $\sim 0.2$ in certain flavor physics models. We
also discuss both upper and lower bounds on the flavor symmetry breaking
scale. We conclude the Letter in \S 5.

\newsec {Dimension Five $B$-Violating Operators}

We begin by summarizing the properties of the dimension five $B$
violating operators in supersymmetry, following the notation of
\ref\hmy{J. Hisano, H. Murayama and T. Yanagida, \np{402}{1993}{46}}.
These operators are either composed entirely out of the weak doublet
superfields $Q$ and $L$ and take the form $\db=QQQL$, or else involve
only the weak singlet superfields, $\sb=U^cU^cD^cE^c$.  Because of
antisymmetrization in color, at least two families must be involved; to
avoid suppression by small mixing angles, we need only consider the
quark superfields from the lightest two families.  The singlet operators
$\sb$ make a much smaller
contribution to proton decay and we will ignore them,
since they require at least one power
each of the $c$ quark and $\tau$ lepton
Yukawa couplings to convert the right-handed
charm quark superfield into a lighter flavor.  The $\db$ operators have
two types of flavor structure for the quarks:
\eqn\lops{\eqalign{
\pdi &= (Q_1 Q_1)(Q_2 L_i)=
2\epsilon_{abc}(U_aD'_b)(C_cE_i - S'_cN_i)
\ ,\cr
\pdii &= (Q_2 Q_2)(Q_1 L_i)=
2\epsilon_{abc}(C_aS'_b)(U_cE_i - D'_cN_i)
\ .\cr}}
In the above equation we take $Q$ to be the $SU(2)$ eigenstates
$Q=(U,D') = (U,VD)$, where $U$ and $D$ are mass eigenstates and $V$ is
the CKM matrix.  The bracketed quantities are $SU(2)$ singlets, and $i$
refers to the lepton family, with $i=1,2$ for the charged leptons $E$ or
$1,2,3$ for the neutrinos $N$. Operators of the form $(Q_1Q_2)(Q_2 L_i)$
and $(Q_1Q_2)(Q_1 L_i)$ can be Fierz rearranged into the above forms.

The operators \lops\ may be dressed with gauginos to yield the
four-fermion operators in the low energy, nonsupersymmetric theory that are
relevant for proton decay. There are five types of operators relevant for
proton decay; two that result in a positively charged lepton in the final
state:
\eqn\fourq{
\CO_{1i} = (su)(ue_i)\qquad
\CO_{2i} = (du)(ue_i) \ ,}
where $i=1,2$, and three operators giving rise to antineutrinos in the final
state:
\eqn\forqii{
\CO_{3i} = (du)(s\nu_i)\qquad
\CO_{4i} = (su)(d\nu_i)\qquad
\CO_{5i} = (du)(d\nu_i)
\ ,}
where $i=1,2,3$. Our notation is such that
$$(su)(ue_i)\equiv\epsilon_{abc}\epsilon_{\alpha\beta}\epsilon_{\gamma\delta}
(s_L^{a\alpha}u_L^{b\beta})(u_L^{c\gamma} e_{Li}^{\delta})\ ,$$ etc.
Assuming that the squarks are nearly degenerate, gluinos do not
contribute, while wino dressing at one loop yields
\eqn\dress{\pdi \to
{\alpha_2\over \pi}\sum_{n=1}^5 a_{ni} \CO_{ni}\ ,\qquad \pdii\to
{\alpha_2\over \pi}\sum_{n=1}^5 b_{ni} \CO_{ni}\ ,}
where $\alpha_2 = \alpha/\sin^2\theta_w$ and
\eqn\coeff{\eqalign{
a_{1i} &= -\cc[f(c,d') + f(\nu_i,d')]\cr
a_{2i} &=-\tc a_{1i}\cr
a_{3i} &= -\cos^2\theta_c[f(c,e_i)+f(u,d')]\cr
a_{4i} &=-\tan^2\theta_c a_{3i}\cr
a_{5i} &=-\tan\theta_c a_{3i}\cr}
\qquad
\eqalign{
b_{1i} &=0\cr
b_{2i} &=0\cr
b_{3i} &=-\cc\sc[f(c,e_i) + f(c,d')]\cr
b_{4i} &= b_{3i}\cr
b_{5i} &= -\tan\theta_c b_{3i}\ .}}
The function $f$ is given by \ref\na{P.~Nath and R.~Arnowitt,
\physrev{D26}{1982}{287}}
\eqn\fdef{
 f(u,d)\equiv
{\mw\over\mus^2-\mds^2}\(
{\mus^2\over\mus^2-\mw^2}\ln{\mus^2\over\mw^2}-
 {\mds^2\over\mds^2-\mw^2}\ln{\mds^2\over\mw^2}\)\ .}
For degenerate squarks with mass $m_{\tilde q}\gg \mw$,  $f$ is simply given by
$f\simeq\mw/m^2_{\tilde q}$.

   We
can quantify the effects of the operators \lops\ by parametrizing their
strength in the superpotential $W$ in terms of the modified Planck scale
\eqn\planck{M_p^{\star}\equiv {M_p\over \sqrt{8\pi}}  = 2.4\times 10^{18}\
\GeV}
and the dimensionless coupling constants $g_{1,2}$:
\eqn\wfive{W_5 = {1\over M_p^{\star}}\sum_{i=1}^3 \[
g_{1i}\pdi + g_{2i}\pdii\]\ .}
In terms of these $g$ coefficients one finds the widths \hmy
\eqn\pwidths{\eqalign{
\Gamma(p\to K^0\ell^+_i) &= \({\CA\beta\alpha_2\cos\theta_c\over \pi
M^{\star}_p }\)^2{(m_p^2-m_K^2)^2\over 32 \pi m_p^3f_{\pi}^2}
\Bigl\vert g_{1i} \kappa_1[f(c,d') + f(\nu_i,d')]\Bigr\vert^2\cr
\Gamma(p\to K^+\bar\nu_i) &= \({\CA\beta\alpha_2\cos\theta_c\over \pi
M^{\star}_p }\)^2{(m_p^2-m_K^2)^2\over 32 \pi m_p^3f_{\pi}^2}\cr
&\times\Bigl\vert
 g_{1i} \kappa_2
\cos\theta_c[f(c,e_i) + f(u,d')] +g_{2i} \kappa_3
\sin\theta_c[f(c,e_i) + f(c,d')]\Bigr\vert^2\cr
\Gamma(n\to K^0\bar\nu_i) &= \({\CA\beta\alpha_2\cos\theta_c\over \pi
M^{\star}_p }\)^2{(m_n^2-m_K^2)^2\over 32 \pi m_n^3f_{\pi}^2}\cr
&\times\Bigl\vert
 g_{1i} \kappa_2
\cos\theta_c[f(c,e_i) + f(u,d')] +g_{2i} \kappa_4
\sin\theta_c[f(c,e_i) + f(c,d')]\Bigr\vert^2\cr}}
where $\beta$ is an unknown strong matrix element, estimated to be $\sim
10^{-2}\ \GeV^3$, while the $\kappa$'s are coefficients computable from
the QCD chiral Lagrangian in terms of the axial current matrix elements
$D$ and $F$ \ref\mine{M.~Claudson, M.B.~Wise and L.J.~Hall,
\np{195}{1982}{297}}, \hmy:
\eqn\kapval{\eqalign{
\kappa_1 &\equiv 1 - {m_p\over m_{\Lambda}}(D-F) =  0.70 \cr
\kappa_2 &\equiv 1 + {m_p\over 3 m_{\Lambda}}(D+3F) = 1.6 \cr
\kappa_3 &\equiv 1 + {m_p\over m_{\Lambda}}(D+F) = 2.0\ .\cr
\kappa_4 &\equiv 2 + {2m_p\over m_{\Lambda}}F = 2.7\ .\cr}}
The above formulae include one loop scaling effects due to the gauge
interactions from
$M_p^{\star}$ down to 1 GeV, which give an enhancement of $\CA \simeq 10.5$ in
the amplitude.
Again, note that $\pdi$ makes the sole contribution to $p\to K^0\ell_i^+$;
furthermore, the contribution of $\pdii$ to $p\to K^+\bar\nu_i$ is Cabbibo
suppressed.  Since $\kappa_2>\kappa_1$ the most stringent experimental
limits on the $g$ couplings come from \ref\pdb{K.~Hikasa {\it et al}\/.,
Particle Data Group, Phys. Rev. D45 (1992), II.25}
$$\Gamma(p\to K^+\bar\nu) < (1.0\times 10^{32}\,yr)^{-1},$$
$$\Gamma(n\to K^0\bar\nu) < (8.6\times 10^{31}\,yr)^{-1},$$
which yield
\eqn\plife{\eqalign{
\sqrt{\sum_i \vert g_{1i}\vert^2} &< 
			3.6\times 10^{-8} \times
\({1\,{\rm TeV}^{-1}\over f(c,d')+f(u,\ell_i)}\)\({ 0.01\,{\rm
GeV}^3\over\beta}\)
\ ,\cr
 \sqrt{\sum_i \vert g_{2i}\vert^2} &< 
			1.0\times 10^{-7} \times
\({1\,{\rm TeV}^{-1}\over f(c,d')+f(c,\ell_i)}\)\({ 0.01\,{\rm
GeV}^3\over\beta}\)
\ .\cr }}

The charged lepton modes are in general smaller than similar neutrino modes
because (i) $g_{2i}$ only contributes to the neutrino modes, and (ii)
the $g_{1i}$ contributions are larger for
neutrino modes than for charged lepton
modes since the ratio of the amplitudes contains a factor of
$\kappa_1/\kappa_2= 0.4$.  As we discuss in the
following section, in conventional GUT models $g_{1i}\ll g_{2i}$
and the charged lepton modes are invisible.

\newsec {Predictions from GUT models}

Consider a generic GUT theory where the only breaking of the chiral
flavor symmetries is due to Yukawa interactions with the same general
size and texture of the Yukawa couplings in the standard model at low
energy, or smaller.  This includes GUTs  with
non-minimal Higgs field content, such as the Georgi-Jarlskog
model \ref\gj{H.~Georgi and C.~Jarlskog, \pl{86}{1979}{297}}, for example. It
follows that all of the color triplet scalars will
have couplings of the form $Q\tilde Y Q$ and $Q\tilde Y' L$, where
$\tilde Y \ltap Y_U$ and $\tilde Y'\ltap Y_D$  (or similarly with the $U$ and
$D$ subscripts reversed, in
the case of a ``flipped'' charge  embedding).
In such theories the  $\db_{\{1,i\}}$ operators are generated with strength
$\tilde Y_{11}$ ($ \tilde Y'{}^*_{2i}$) while the  $\db_{\{2,i\}}$ operators
are generated with strength $\tilde Y_{22}$  ($ \tilde Y'{}^*_{1i}$).
As such,
 one sees that the $\pdi$ operators are suppressed relative to $\pdii$ by
$\sim \sqrt{m_u/m_c}$.  Therefore the charged lepton decay modes of the proton
are unlikely to be seen.

For example, in minimal $SU(5)$ \hmy, colored Higgs exchange generates
operators
of the form
\eqn\minsuf{{1\over {2M_{H_c}}} y_{u_i} y_{d_k} V^*_{jk}
			(Q_i Q_i)(Q_j L_k)\ .}
Operators involving third generation quark fields can be at most
comparable to those with first two generation fields only in \minsuf, 
only if one takes extremal values for CKM
angles in the range  presently allowed by experiments. Therefore hereafter we
only discuss operators involving quark fields from the first two generations,
the relevant terms being:
\eqn\minsu{W_5 \sim {1\over 2M_{H^c}}\[ y_u y_{d_i} V^*_{2i} \pdi
+ y_c y_{d_i} V^*_{1i} \pdii \]. }
Evidently only $\db_{\{2,2\}}$ is relevant for proton decay since the other
operators are suppressed by an additional power of $\sin\theta_c$, $y_u/y_c$,
or both.  Since the $p\to K^0\ell^+$ decay can only proceed through
 $\db_{\{1,1\}}$ and  $\db_{\{1,2\}}$, one finds that
 $BR(K^0 \ell^+)/BR(K^+\bar\nu)$ contains a suppression factor of
 $( y_u\kappa_1/y_c\kappa_3\sin^2\theta_c)^2\sim 6\times 10^{-4}$.
 In the more realistic GUT example of ref.
\ref\ardhs{G.~Anderson, S.~Raby, S.~Dimopoulos, L.J.~Hall, and
G.D.~Starkman, \physrev{D49}{1994}{3660}.}, the effective superpotential
for proton decay is \ref\ahmr{A.~Antaramian, L.J.~Hall,
H.~Murayama, and A.~Rasin, in preparation.}
\eqn\soph{W_5 \sim
	{y_c y_s \over 2 M_{H^c}} {50 \gamma y_s \over y_c}
	\sqrt{y_d \over y_s} \[
	\delta {y_u\over y_c} \db_{\{1,1\}}
	+ \delta \sqrt{y_u \over y_c } \db_{\{2,1\}}
	+ \sqrt{y_u \over y_c}  \db_{\{1,2\}}
	+ \db_{\{2,2\}}\] , }
where the coefficients $\gamma \sim 1$, $\delta \sim 0.5$ are combinations of
model dependent Clebsch--Gordon factors.
The suppression of charged lepton decay modes is an order of magnitude
less  severe than in minimal $SU(5)$.
Nevertheless, one finds from
eqs. \wfive-\kapval\ that for this model
\eqn\hallmod{\eqalign{
{BR(p\to K^0\mu^+)\over BR(p\to K^+\bar\nu)} &\simeq 8\times 10^{-3}\ ,\cr
{BR(p\to K^0e^+)\over BR(p\to K^+\bar\nu)} & \simeq 6\times 10^{-6}\ ,\cr}
\qquad {\rm (model\ of\ ref.\ \ardhs)}}
and so $p\to K^0\mu^+$ remains unlikely to be
seen, while $p\to K^0e^+$ is still certainly undetectable.

In a ``flipped'' GUT the roles played by the up and down Yukawa
couplings is effectively reversed, and the quark doublet $Q=(U,D')$ is
replaced by $Q' = (U',D) = V^{\dagger} Q$.  Furthermore, the lepton doublet no
longer involves mass eigenstates for the charged leptons, and is given  by
$L' = (\nu, \ell') = (\nu, V_{\ell} \ell)$, where $V_{\ell}$ is the lepton
analogue of the CKM matrix $V$.  Thus Eq. \minsuf\ is replaced by
\eqn\fsuf{{1\over {2M_{H_c}}} y_{d_i} y_{u_k} V_{kj}
(Q'_m Q'_n)(Q'_j L'_k)\ .}
With these interactions one finds the dominant contribution to $p\to
K^+\nu_\tau$ to be of the form $y_d y_t V_{ts} (Q'_1Q'_1)(Q'_2L'_3)$, without
the further
accompanying
$V_{ts}$ suppression found in the minimal $SU(5)$ example.  If one assumes
that $V_{\ell}$ is close to the unit matrix, then one
 finds that $p\to K^0\mu^+$ generated by $y_s y_c V_{cd} (Q'_2Q'_2)(Q'_1L'_2)$
dominates the
charged lepton modes, but at a rate smaller than $p\to K^+\bar\nu$ by a
factor of
\eqn\flippo{
{BR(p\to K^0\ell^+)\over BR(p\to K^+\bar\nu)}\sim
\({m_sm_cV_{cd}V_{us}^* \kappa_1\over m_dm_tV_{ts} \kappa_2}\)^2\simeq 8\times
10^{-3}\qquad ({\rm flipped}\ SU(5),\ V_{\ell}\sim  1)}
 so that
it is still invisible.%
\foot{
If $V_{\ell}$ has large off-diagonal elements
({\it e.g.}\/ $(V_{\ell})_{\nu_\tau e} \simeq 1$),
then the above ratio can
be as large as $\CO(1)$, and the theory is an example of how first family
chiral symmetries being badly broken by new flavor physics; in this case
the new flavor physics is in the lepton sector.
}
Furthermore, in
flipped $SU(5)$ it is possible to make the parameter $M_{H^c}$ be
effectively much larger than the GUT scale, or infinite
\ref\aehn{I.~Antoniadis, J.~Ellis, J.S.~Hagelin, and D.V.~Nanopoulos,
\pl{194}{1987}{231}}\ so that all proton decay modes are invisible.

It follows then that observing the decay $p\to K^0 \ell^+$  (and $p\to K^0e^+$
most strikingly)  will be a  signal for flavor physics with a structure
different than the low energy Yukawa couplings.  It is easy to invent
an unnatural example of such a model --- for example, consider $SU(5)$
with an extra pair of superfields $\phi$, $\bar \phi$ transforming as
$5$, $\bar 5$ respectively, with a GUT scale mass and a coupling to
fermions of the form
\eqn\toy{f^u_{ij} \phi 10_i 10_j + f^d_{ij} \bar \phi  10_i\bar{5}_j \ ,}
where the $f$ couplings have no hierarchical structure and are small
enough to be consistent with experimental bounds on the proton lifetime.
In such a model the operators $\pdi$ would be generated with the same
strength as $\pdii$, and one might see $p\to K^0 \ell^+$ --- precisely
because the $f$ matrices break the chiral symmetries for the light
families in a different
(bigger) way than do the Yukawa couplings.  This example is illustrative
but not interesting, since it renders the structure of the Yukawa
coupling inexplicable instead of merely mysterious.  In the next section
we point out how a model with softly broken flavor symmetries which
explain the structure of the Yukawa couplings can also give rise to
detectable $p\to K^0\ell^+$ decay.

\newsec{Proton Decay from Planck Scale Physics}

For simplicity we will first consider the models without grand
unification with a flavor symmetry $G_f$ broken by the expectation
values of some fields $X$.  The flavor symmetry breaking is assumed to
be transmitted to the quarks and leptons by means of particles with mass
$M$ so that the low energy Yukawa couplings are constructed out of
powers of $\epsilon\equiv\vev{X}/M$ with a texture dictated by $G_f$
symmetry \ref\frog{C.D.~Froggatt and H.B.~Nielsen, \np{147}{1979}{277}}.
Such models can explain the gross
hierarchical features of the quark and lepton mass matrices, such as in
the recent examples \nref\seiberg{M.~Leurer, Y.~Nir, and N.~Seiberg,
\np{398}{1993}{319}; RU-93-43, hep-ph/9310320; Y.~Nir and N.~Seiberg,
\pl{309}{1993}{337}.}\nref\sp{P.~Pouliot and
N.~Seiberg, \pl{318}{1993}{169}.}\nref\ks{D.B.~Kaplan and M.~Schmaltz,
\physrev{D49}{1994}{3741}}\nref\ir{L.~Ibanez and G.G.~Ross,
OUTP-9403, hep-ph/9403338.}\refs{\seiberg-\ir}.
It is preferred, moreover, that a theory which explains the quark mass
matrix also explains at the same time why the squarks do not give rise
to large FCNC. Note that one cannot ensure squark degeneracy
 when $G_f$ is only an abelian symmetry.
It is possible to remedy this with a clever choice of abelian symmetries, as
in \seiberg, where nondegenerate squarks have mass matrices that align
with the quark masses; a more automatic solution is to enforce
degeneracy by having a nonabelian symmetry $G_f$ throughout the theory,
with at least the first
two families in an irreducible representation \ref\dine{M.~Dine,
R.~Leigh, and A.~Kagan, \physrev{D48}{1993}{2214}.}, \refs{\sp,
\ks}

We will take the point of view that an explanation for the lack
of FCNC    and for
the structure of the quark and lepton mass matrices  mandates some sort of
flavor symmetry, most likely nonabelian,  at short
distances.  What we point out here is that the flavor symmetries can
also control other flavor changing effects, particularly in proton
decay.  In particular, the dimension five operators \wfive\ may be
forbidden by $G_f$ symmetry, while the dimension six operator
\eqn\dsix{{g'\over( M^{\star}_p)^2} QQQLX}
is allowed, in which case one might expect Planck scale physics to
generate it with some $\CO(1)$ coupling $g'$. Such is the case with the
nonabelian discrete symmetry $\Delta(75)$ discussed in ref. \ks; there
the three families of $Q$ and $L$ transform in the fundamental triplet
representation $T_1$, while $X$ is in the representation $T_2$ and
acquires a $Z_3$ preserving vacuum expectation value $\vev{X} =
(\Lambda,\Lambda,\Lambda)$. One finds that $\Delta(75)$ symmetry forbids
the dimension five operators $QQQL$, while there are several different
invariants of the form \dsix. Below the $\Delta(75)$ symmetry breaking
scale $\Lambda$, this results in an effective superpotential of the form
\wfive\ with coefficients
\eqn\gval{g_{1i} \simeq g_{2i}\simeq {g'\Lambda\over M^{\star}_p}\ .}
Proton decay through the dimension five operators becomes naturally
suppressed, without being eliminated entirely, as would be the case if
one imposed the anomaly free symmetry ${\rm exp}(2\pi i B/3)$
\ref\ir2{L.~Ibanez and G.G.~Ross, \np{368}{1992}{3}}.
Furthermore, the $\pdi$ operator dominates because of Cabbibo
suppression in the $b$ coefficients of eq. \coeff, and so one finds
\eqn\brats{
BR(p\to K^0\ell^+)/BR(p\to K^+\bar\nu) \simeq (\kappa_1/\kappa_2)^2 = 0.2\ .}
Since $p\to K^+\bar\nu$ is the dominant decay mode, it follows that if
proton decay is seen at all, it will probably be possible to see the
charged lepton decay mode. It should be added that the charged lepton
modes are easier to reconstruct than the neutrino modes in the
\v{C}erenkov detectors, especially in the
$p \to K^0 \ell^+$, $K^0 \to \pi^0 \pi^0 \to \gamma\gamma\gamma\gamma$
final state.

Since the effects of the operator get larger as $\vev{X}$ gets
larger,  there is now
an experimental {\it upper} bound for the scale of $G_f$ symmetry
breaking $\Lambda$:
\eqn\uboundvi{g'\Lambda \ltap 9\times 10^{10} \ \GeV\qquad {\rm (dim.\ 6)},}
with the same uncertainties given in \plife. This bound is weakened if
for a different $G_f$ symmetry the dimension six operator \dsix\ was
forbidden (\eg, by an additional $U(1)$ or discrete flavor symmetry) and
 only a dimension 7 operator was allowed; then one would
find
\eqn\uboundvii{\sqrt {g'} \Lambda \ltap 5\times 10^{14}\ \GeV\qquad {\rm (dim.\
 7)}.}

If $g'\Lambda$ or $\sqrt{g'}\Lambda$ are near the upper bounds \uboundvi,
\uboundvii, then we might hope to see $p\to
K^0\ell^+$ at Super-Kamiokande; so what can be said about this scale $\Lambda$?
One can derive a crude lower bound on this scale by considering the running
of the gauge coupling constants.  Models of the sort considered here
have the quark and
lepton masses arise from a ``see-saw'' mechanism, whereby the light
particles mix with Dirac counterparts of mass $M$ through the Higgs
doublets and the symmetry breaking field $X$.  Typically $\Lambda/M\sim
10^{-1}$ in order to explain the size of the observed mixing angles and
mass ratios.
 It follows that there is a fermion contribution to
the  $\beta$ functions which is at least three times the usual one,
and this typically renders the gauge forces asymptotically unfree above
the scale $M$; one can then derive a lower bound on $M$ --- and hence
$\Lambda$ --- by requiring that all Landau poles be above the Planck
scale.  For example, in the un-unified standard model with an additional
three Dirac families with mass $M$, the one-loop $\beta$ function for
hypercharge leads to a Landau pole below the scale $M_p^{\star}$ if $M$
is below $6\times 10^{13}$ GeV.  Thus for a flavor symmetry breaking
scale $\Lambda\simeq M/20$, as in ref. \ks, avoiding a Landau pole
implies the lower bound
\eqn\lowb{\Lambda \gtap 3\times 10^{12}\,{\rm GeV}\ ,}
a value that is easily consistent with the upper bound \uboundvii, but not
with \uboundvi\ assuming  that $g'=\CO(1)$.  This suggests that proton decay
operators
in such models must be protected by flavor symmetries,
at least through dimension six.
Dimension seven operators of the form \dsix, with an extra factor of
$X/M^{\star}_p$ are allowed so long as \lowb\ and \uboundvii\ are satisfied.
 The Super-Kamiokande experiment is expected to
improve bounds on the proton decay by a factor of $\sim 30$, which would
be sensitive to dimension seven operators with $\Lambda$ lower than the bound
 \uboundvii\ by a factor of $\sim 5-6$.  Thus it is possible, but not
necessary, that the effects of dimension seven operators could be detected if
close to the bound \uboundvii.
 We note in passing  that at the bound \uboundvii,
$\Lambda\sim 5\times 10^{14}$ GeV, then the
ratio $\Lambda/M_{GUT} \sim 1/20$  could conveniently provide the small
parameter  used in constructing the fermion hierarchy.

If the gauge groups are enlarged or unified below $M_p^{\star}$ the
lower bound \lowb\ on $\Lambda$ might be weakened somewhat, leaving
a small window for dimension six operators, which could be detected at
Super-Kamiokande.  However, extended theories typically require many new
matter
fields in the Higgs sector  which can completely cancel the gauge
contribution to the 1-loop $\beta$ function.
Therefore using unification to lower the bound \lowb\ is not easy to do.

\newsec{Conclusions}

We have shown that detecting the decay mode $p\to K^0\ell^+$ would be a
certain signal for new flavor physics at short distances in a
supersymmetric theory.  Such decay modes can be dominant if the
dangerous dimension 5 proton decay operators generated at the Planck
scale are forbidden only by flavor symmetries which allow higher
dimension $B$ violating operators. Unlike other probes for new flavor
symmetries, these effects become stronger as the flavor scale $\Lambda$
becomes $higher$.
In principle such operators could be very small for
$\Lambda\ll M^{\star}_p$. However for most theories generating fermion
masses {\it a la} Froggatt and Nielsen, the loss of asymptotic freedom
at short distances places a lower bound on $\Lambda$.
The bound turns out to be within a factor of
$\sim 10^2$ of the
upper bound on $\Lambda$ from proton decay, suggesting that the
Super-Kamiokande experiment may be able to detect their effects.
In
such theories, proton decay is not itself a signal for a unification of
gauge forces.

It is worth noting in conclusion that in principle nonabelian flavor
symmetries can serve not only to explain the quark and lepton masses,
squark degeneracy, and the suppression of dimension 5 $B$ violating
operators, but may also control the dimension $\le 4$ $B$
and $L$ violating operators.  The flavor symmetry $\Delta(75)$ mentioned
in above and in ref. \ks\ does not eliminate the dimension four $U^cD^cD^c$
operator (being a subgroup of $SU(3)$), but other symmetries could.  In
such a model, $R$-parity could arise as an accidental symmetry of the
renormalizable theory broken only by $M_p^{\star}$ suppressed operators,
making the $LSP$ unstable, but with a very long lifetime.%
\foot{The strongest constraint from proton decay
is $\lambda_{U^c D^c S^c} \lambda_{Q_1 S^c L_{1,2}} < \CO(10^{-26})$.
This suggests $\lambda < \CO(10^{-13})$, roughly the square of first
generation Yukawa couplings. If this were the typical size of
$R$-parity violation, then the $LSP$ would decay near the time
of nucleosynthesis,
threatening light element abundances.
However $R$-parity violation due to other operators is  constrained only
by the cosmological requirement that $\lambda < \CO(10^{-8})$ in order
to preserve the baryon asymmetry \ref\cdeo{B.A.~Campbell, S.~Davidson, J.~Ellis
and K.A.~Olive, \pl{256}{1991}{457}}.  $R$-parity violation at this
level allows the $LSP$ to decay long before nucleosynthesis. The challenge for
flavor model building is to satisfy these two constraints simultaneously.}
\vfill\eject
\centerline{Acknowledgements}

HM is grateful to Aram Antaramian, Lawrence Hall and Andrija Rasin for
discussions, and we both thank the Institute for Theoretical Physics at UC
Santa Barbara for hospitality where part of this work was done under
 NSF  grant  PHY89-04035.
HM was supported in part by the Director, Office of Energy
Research, Office of High Energy and Nuclear Physics, Division of High
Energy Physics of the U.S. Department of Energy under Contract
DE-AC03-76SF00098;  DK was supported in part by DOE grant DOE-ER-40561, NSF
Presidential Young Investigator award \pyidk, and by a grant from the
Sloan Foundation.

\listrefs

\bye